\documentclass{article}
\usepackage{spconf,amsmath,graphicx}
\usepackage{comment}
\usepackage{color}

\usepackage{amsmath}
\usepackage{url}
\usepackage{multirow}
\usepackage[utf8]{inputenc}
\usepackage{CJK}

\usepackage[whole]{bxcjkjatype}
\usepackage{appendix}

\newcommand*{\red}{\textcolor{black}}

\usepackage{bbding}
\usepackage{pifont}
\usepackage{wasysym}
\usepackage{amssymb}
\usepackage{comment}
\usepackage{xcolor}
\usepackage{tablefootnote}
\usepackage{enumitem}

\title{Benchmarking Japanese Speech Recognition on ASR-LLM Setups with Multi-Pass Augmented Generative Error Correction}
\name{Yuka Ko${}^{1}$, Sheng Li${}^{2}$, Chao-Han Huck Yang${}^{3}$, Tatsuya Kawahara${}^{4}$}
\address{${}^{1}$Graduate School of Informatics, Nara Institute of Science and Technology (NAIST), Ikoma, Nara, Japan \\
${}^{2}$National Institute of Information and Communications Technology (NICT), Kyoto, Japan\\
${}^{3}$NVIDIA Research, Taiwan \\
${}^{4}$Kyoto University, Sakyo-ku, Kyoto, Japan 
}
\begin{document}
%
\maketitle
\begin{abstract}
With the strong representational power of large language models (LLMs), generative error correction (GER) for automatic speech recognition (ASR) aims to provide semantic and phonetic refinements to address ASR errors. 
This work explores how LLM-based GER can enhance and expand the capabilities of Japanese language processing, presenting the first GER benchmark for Japanese ASR with 0.9-2.6k text utterances. 
We also introduce a new multi-pass augmented generative error correction (MPA GER) by integrating \textit{multiple system hypotheses} on the input side with corrections from \textit{multiple LLMs} on the output side and then merging them.
To the best of our knowledge, this is the first investigation of the use of LLMs for Japanese GER, which involves second-pass language modeling on the output transcriptions generated by the ASR system (e.g., $N$-best hypotheses). 
Our experiments demonstrated performance improvement in the proposed methods of ASR quality and generalization both in SPREDS-U1-ja and CSJ data.

\end{abstract}

\section{Introduction}
\label{sec:intro}
The automatic generative error correction (GER) task involves using pre-trained models and algorithms (e.g., contextual biasing) to automatically revise and improve the quality of written content, such as documents, articles, or translations. This process typically includes correcting grammar, syntax, style, and coherence errors and adjusting language to meet specific requirements or preferences. Automatic GER tasks aim to streamline the editing process, saving time and effort while enhancing the overall readability and effectiveness of the post-recognition transcript quality.

GER tasks involve \textit{revising} outputs from automatic speech recognition (ASR) systems. When speech is transcribed into text, errors can arise due to variations in accents, background noise, or misinterpretations of words. Recent advanced GER algorithms, such as text-to-text modeling~\cite{yang2023generative} and speech-language representation alignments~\cite{radhakrishnan2023whispering}, address these issues in ASR transcriptions. For instance, GER-based methods correct inaccuracies, enhance grammar and syntax, and ensure that the text accurately represents the spoken content. This refinement process significantly improves the quality and usability of the transcribed text, making it more valuable for applications such as captioning, transcription services, or voice-controlled systems.

\begin{figure}[t]
  \centering
  \includegraphics[width=0.48\textwidth]{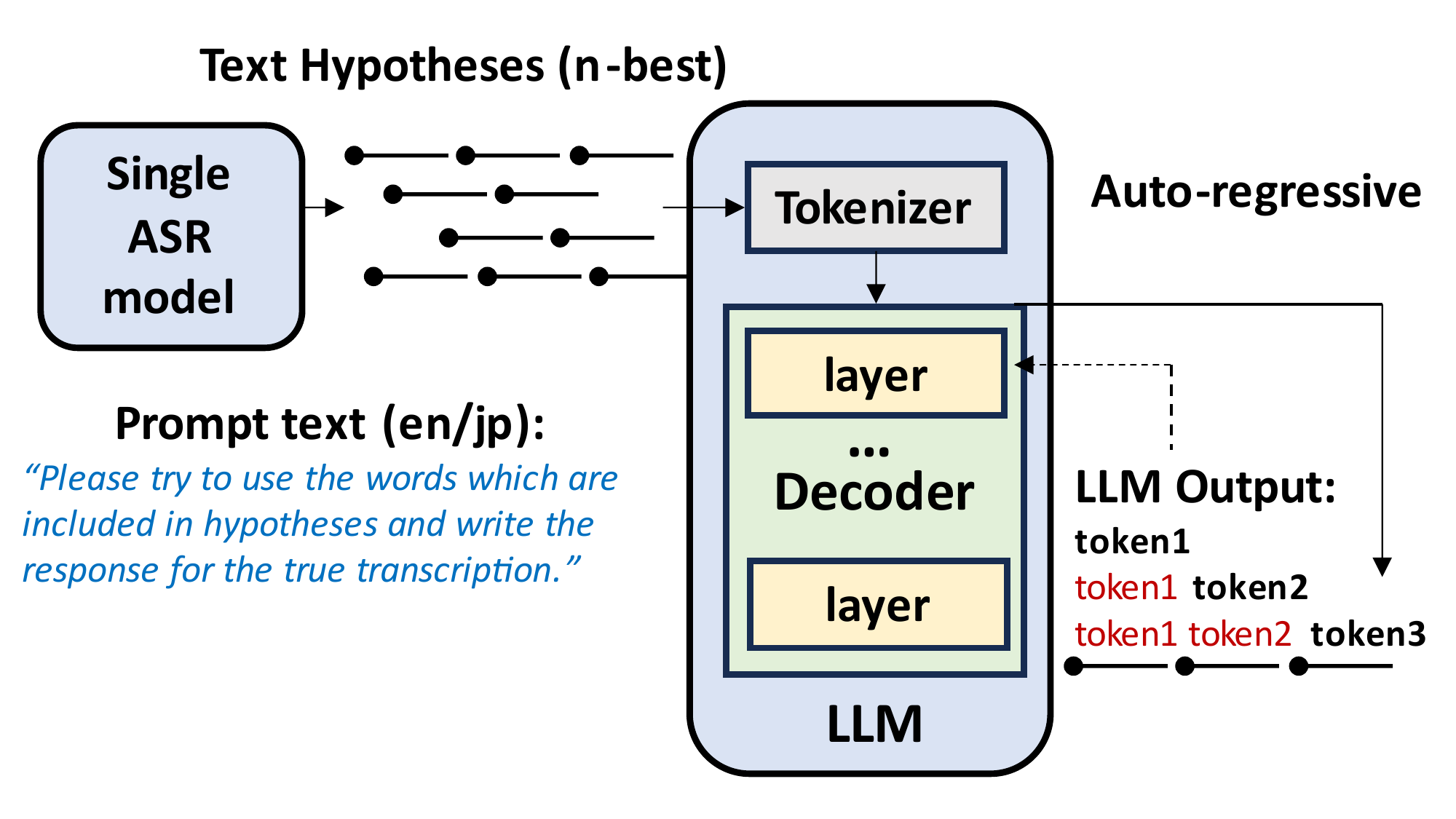}
  \caption{The standard LLM GER method rescoring $N$-best hypotheses.}
  \label{fig:method}
\end{figure}

\begin{figure*}[t]
\setlength{\abovecaptionskip}{9pt}
  \centering
  \includegraphics[width=0.8\textwidth]{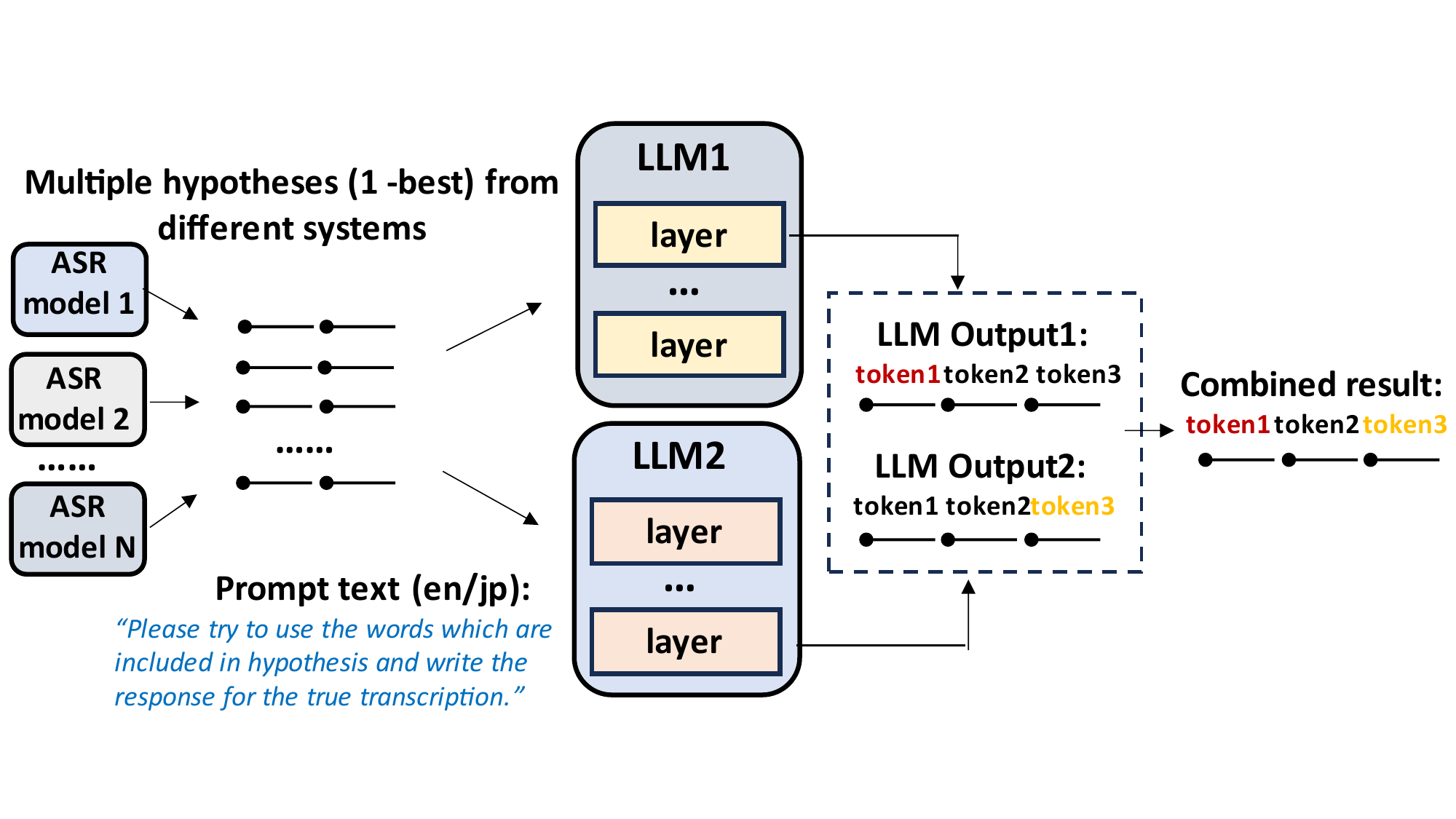}
  \vspace{-30pt}
  \caption{The proposed multi-pass augmented (MPA) GER method combines hypotheses from different ASR and LLM models.}
  \label{fig:method2}
\end{figure*}

In recent years, pretrained big neural network-based language models (LMs) have been used in ASR-related GER tasks. Zhang \emph{et al.} \cite{zhang2019investigation} proposed a spelling corrector based on the transformer \cite{attisall} to reduce the substitution error in Mandarin speech recognition. Zhang \emph{et al.} \cite{zhang2020spelling} improved the BERT \cite{devlin2018BERT} effectiveness in detecting spelling errors with the soft-masking technique as the bridge between the error detector and corrector. Futami \emph{et al.} \cite{futami2020distilling} generated soft labels for ASR training with the BERT distilling knowledge. There are also some works \cite{salazar2019masked,shin2019effective} studying to improve ASR rescoring by BERT. Additionally, BERT  has also been successfully applied in multi-modal studies in vision-language pretraining \cite{lu2019vilBERT,li2019visualBERT,zhou2020unified,li2020oscar} or voice-language pretraining \cite{baevski2020effectiveness,hsu2021huBERT,wang2020curriculum}. 

More recently, \cite{yang2023generative, Chen2023HyPoradiseAO} proposed combining large language models (LLMs) into a speech recognition system. In this work, we extend the correction methods described in \cite{Chen2023HyPoradiseAO} and apply them to Japanese ASR tasks in different settings \red{from $N$-best hypotheses.} Additionally, Li et al. \cite{li24h_interspeech} investigated knowledge transfer when fine-tuning a multilingual LLM to correct 1-best hypothesis errors generated by different speech foundation models across multiple languages.
%
We introduce a new multi-pass augmented generative error correction (MPA GER) by integrating \textit{multiple system hypotheses} on the input side with corrections from \textit{multiple LLMs} on the output side and then merging them. 
Moreover, We will also address the issue with a fairly low \red{Character Error Rate (CER)} for which the conventional \red{Recognizer Output Voting Error Reduction (ROVER) \cite{fiscus1997post}} and previous LLM GER no longer provide enhancements.
Our results show that MPA GER shows improved performance on ASR quality and generalization.

\section{Related Work}
\subsection{LMs for ASR task}

Using an LM in ASR improves speech recognition performance. Generally, ASR combined with LM has two strategies: first-pass decoding and second-pass rescoring. 

In ASR, the task was formulated as a noisy channel model using the Bayes rule \( P(W|X) = P(X|W)P(W) \), where \(X\) is the speech signal and \(W\) is the corresponding text. The two distributions of \(P(X|W)\) and \(P(W)\) were named acoustic and language models, respectively. The LM was trained separately on the source text and only used for decoding \cite{jelinek1976continuous}. WFST-based decoding compiles n-gram LMs into the decoding graph for efficient first-pass decoding \cite{mohri2008speech}. Incorporating larger n-gram LMs made the decoding graph explode, and researchers changed the compiling improved algorithm \cite{dixon12_interspeech} or used second-pass rescoring in both offline and on-the-fly settings \cite{ljolje1999efficient,sak2010fly} instead. The Bayesian formulation still made sense in the hybrid DNN-HMM model era. The scores could be interpreted as pseudo-likelihoods by subtracting an appropriate prior, so the same decoding/rescoring framework carried over.

For End-to-End models, they directly estimate \(P(W|X)\). We still can combine using LMs in the first-pass decoding (e.g., shallow fusion, cold fusion, etc. \cite{chorowski2016better,sriram2017cold}). Similar to the second-pass rescoring, hypotheses can be obtained using beam search on an ASR model and re-rank with an externally trained LM. A two-pass E2E ASR model was proposed with an encoder shared between a streaming RNN-T model and a full-context LAS decoder \cite{sainath2019twopass}. There are also some works \cite{zhang2019investigation,zhang2020spelling,futami2020distilling,salazar2019masked,shin2019effective} studying to improve ASR rescoring by the transformer and BERT.

\section{\red{Improving generative error correction using LLMs for ASR system}}

%
%
\subsection{LLM GER with $N$-best hypotheses in a single system}

Inspired by \cite{Chen2023HyPoradiseAO}, we establish the LLM GER with single $N$-best hypotheses for Japanese ASR systems and conduct experiments on a dataset with a relatively high recognition errors in order to confirm its effectiveness in Japanese ASR. The system configuration is shown in Fig.\ref{fig:method}.

\subsection{Multi-pass augmented (MPA) GER}
\red{Chen \emph{et al.} \cite{Chen2023HyPoradiseAO} mentioned that the previous LLM GER failed to yield improvements when the WER was low.
There was less room to correct recognition errors with the previous standard LLM GER. 
When the ASR error rate is low, the impact of the errors or hallucinations generated by the LLM may be greater than the impact of ASR error correction.
It may result in a more significant overall error for the LLM GER.
We expect this problem to occur also when applying the previous standard LLM GER in a low CER Japanese ASR task. 
In this study, we propose a multi-pass augmented (MPA) GER approach. This approach, which involves combining hypotheses from different ASR and LLM models, has the potential to significantly enhance LLM GER and make a substantial impact in the field. 
We integrate multiple system hypotheses on the input side and apply corrections from several LLMs on the output side. 
This approach culminates in a merging process akin to a multi-pass combination. 
} 

\red{For MPA GER, we use the ROVER to combine the results with rescoring. 
The ROVER is one of the most efficient among traditional methods for reducing speech recognition errors. 
Its underlying principle involves constructing multiple speech recognizers with comparable performance that operate independently and then vote on their recognition results to lower the overall error rate. 
Our proposed MPA GER utilizes both the advantages of conventional ROVER methods and current LLM approaches. 
In this study, we propose two types of MPA GER: (1) $1$-best hypotheses combination from $N$-systems, (2) $ N$-best hypotheses combination from single system. 
Fig. \ref{fig:method2} shows the overview of MPA GER in $1$-best hypotheses combination from $N$-systems. 
}

\begin{table}[t]\footnotesize
\begin{center}
    \caption{Settings for LLM GER and MPA GER experiments}

    \label{tab:experiment_parameter_setting_llm}
    \begin{tabular}{l|c|c}
    \hline
    Dataset    & SPREDS-U1-ja & CSJ \\ \hline
        LLM & Elyza-7b & Elyza-7b (LLM1) \\
            &                      & Qwen1.5-7b (LLM2) \\ \hline
        Machine & V100 & A6000 \\\hline
        LLM test set & 100 utt.& Eval$_1$ (424) \\ 
                     &            & Eval$_2$ (424) \\
                     &            & Eval$_3$ (424) \\    \hline
        LLM train set & 900 utt. & Eval$_1$+Eval$_2$+Eval$_3$ (2677) \\ 
        & & \red{(except LLM test set)} \\ \hline
        Training bit size & \multicolumn{2}{c}{8-bit} \\\hline
        Epoch & 15 & 20 (Elyza-7b) \\
              &    & 10 (Qwen1.5-7b) \\\hline
        Learning rate & \multicolumn{2}{c}{1e-4} \\\hline
        LoRA rank & \multicolumn{2}{c}{4} \\ \hline
        Prompt Language & English & English-Japanese \\ 
    \hline
    \end{tabular}
\end{center}
\end{table}

\section{Experimental Setup}
There are two sets of experiments in total: one with small data (SPREDS-U1-ja) to verify the basic algorithm (LLM GER) at high error rates and the other larger one (CSJ) with low error rates to validate our improved algorithm (MPA GER). The experimental settings are as follows. Table \ref{tab:experiment_parameter_setting_llm} shows the experiment settings for LLM GER and MPA GER.

\begin{description}[leftmargin=5pt]
 
 \item[LLM models] The experiment employs the Japanese LLM (ELYZA-7b) model\footnote{huggingface.co/elyza/ELYZA-japanese-Llama-2-7b} and Multilingual LLM (Qwen1.5-7b) model\footnote{https://huggingface.co/Qwen/Qwen1.5-7B}. \footnote{\red{In this paper, we refer to the models fine-tuned by these LLMs as Elyza and Qwen1.5, respectively.}}

 \item[Dataset] We use two Japanese datasets for GER. (1) SPREDS-U1-ja \footnote{ast-astrec.nict.go.jp/en/release/SPREDS-U1}. Of these, 900 Japanese sentences are used for fine-tuning process, and a separate set of 100 sentences are used for evaluation. 
(2) the ``Corpus of Spontaneous Japanese (CSJ)" \cite{csj}. \red{For training ASR models, } we use approximately 600 hours of lecture recordings as the training set (Train APS+SPS (academic+simulated)) according to \cite{csj-kaldi, kanda, mitsu2017, CSJ-bench}. Three official evaluation sets (Eval$_1$, Eval$_2$, and Eval$_3$), each containing ten lecture recordings \cite{CSJ-bench}, are used to evaluate the speech recognition results. 
\red{For the LLM test set, we extracted each 424 utterances from each Eval$_1$ (1272), Eval$_2$ (1292), and Eval$_3$ (1385) prepared by the ESPnet2 \cite{watanabe2018espnet} CSJ recipe \footnote{https://github.com/espnet/espnet/tree/master/egs2/csj/asr1}. The left Eval$_1$, Eval$_2$, and Eval$_3$ not included in the LLM test set were gathered and used as our LLM train set (2677 utterances in total). }

 \item[Preparing ASR models] 
 \red{
(1) For the SPREDS-U1-ja GER experiments, we prepared the open-source pre-trained ASR models, including Whisper-Large-v3, Meta MMS model, CMU OWSM v3.1 model, and Whisper-Large's earlier versions (v1 and v2). 
(2) For the CSJ GER experiments, we used the Whisper models and the Conformer-based model after applying fine-tuning for Whisper (Whisper-Large-v2, Whisper-Medium, and Whisper-Small) and training Conformer from scratch by CSJ data.
We applied whisper fine-tuning and conformer training from scratch using CSJ training data, following default settings in ESPnet2 \cite{watanabe2018espnet} CSJ recipe \footnotemark[5] unless otherwise mentioned. 
We did not apply speed perturbation for data augmentation. We fine-tuned all whisper models in 3 epochs with a learning rate of 1e-5. 
We trained the Conformer-based model in 100 max epochs in 4 early-stop patients. 
The training is performed on 1 NVIDIA Tesla A6000 GPU. 
}

\item[Training LLM GER models]
The implementation of our LLM GER models is based on the repository\footnote{https://github.com/Hypotheses-Paradise/Hypo2Trans} for GER task \cite{yang2023generative}. 
In SPREDS-U1-ja, the training is performed on a NVIDIA Tesla V100 GPU using 8-bit training. The hyperparameters for finetuning are 15 epochs, learning rate 1e-4, batch size 64, and LoRA rank 4. 
We used an English prompt in the repository\footnote{https://github.com/Hypotheses-Paradise/Hypo2Trans/blob/main/H2T-LoRA/templates/H2T-LoRA.json} for training and testing without any change. 
In CSJ data, the training is performed on an NVIDIA Tesla A6000 GPU using 8-bit training. 
The hyperparameters for finetuning are 20 epochs in Japanese-Llama-2-7b, 10 epochs in Qwen1.5-7b, learning rate 1e-4, batch size 256, and LoRA rank 4. 
We used an English-Japanese prompt based on automatic translation following the previous default English prompt \footnote{In our pilot study for CSJ data, we got roughly lower loss and lower CER trends in Elyza and Qwen1.5 using the English-Japanese prompt.}. 

\item[For system combination in ROVER]
\red{In CSJ data, we applied ROVER \cite{fiscus1997post} when getting a combined result from multiple LLM outputs for the proposed MPA GER in Fig. \ref{fig:method2}. 
We also applied ROVER directly for multiple ASR hypotheses as the baselines compared with the LLM GER methods. 
We tried two types of MPA GER: (1) $1$-best hypotheses combination from $N$-systems, (2) $N$-best hypotheses combination from a single system. 
In (1), we tried $3$-systems and $4$-systems for $1$-best hypotheses combination. 
In (2), we tried $3$-best hypotheses combination from a single system. 
}

\item[GER Evaluation]
We used NIST-SCTK to evaluate the \red{CER}. 
In CSJ, before the CER calculation, we converted full-width characters to half-width characters and removed punctuation marks from the hypothesis and reference. 
\end{description}

\begin{table}[t] 
\small
\begin{center}
{
\caption{Evaluation of LLM GER with \red{CER [\%]} on Whisper large v3 results of SPREDS-U1-ja}
\label{tab:music}
\begin{tabular}{c|c}
\hline
Before LLM GER correction & After correction  \\
\hline
12.91  & \textbf{7.77}  \\
\hline
\end{tabular}
}
\end{center}
\end{table}

\begin{table}[t]\footnotesize
\centering
\caption{Evaluation of \red{LLM GER} by combining different systems of MMS, OWSM, and Whisper large (WPL) with \red{CER [\%]} on SPREDS-U1-ja}
\begin{tabular}{c|c|c|c|c|c}
\hline
  \multicolumn{5}{c|}{Before proposed MPA GER} & After \\
  \hline
 MMS & OWSM v3.1 & WPL v1 & WPL v2 & WPL v3 & \red{LLM GER} \\
\hline
 32.41 & 10.18 & 9.29 & 9.20 & 12.91  &  \textbf{7.07}\\
\hline
\end{tabular}
\label{table:performance_changes-small}
\end{table}

\section{Experimental Results and Analyses}
\subsection{SPREDS-U1-ja: \red{LLM GER in $N$-best hypotheses and $N$-system combination}}
Table \ref{tab:music} shows the results of the experiments using $N$-best hypotheses. Experiments show LLM GER can effectively improve the performance of the original system output, reflected in the CER. For the $N$-best GER experiment, we conduct the paired sample $t$-test on the results from the best model (Epoch 11). The paired sample $t$-test suggests that the improvement in CER is statistically significant.
Another discovery is switching to the Japanese language prompt for Japanese LLM GER; in the SPREDS-U1-ja dataset, CER can be further reduced from 7.77\% to 7.63\%.
Table \ref{table:performance_changes-small} shows the performance changes before and after using the LLM-based system combination. It has the current best result of GER correction. The leap forward in performance is due to the diversity of different system hypotheses. 
Both LLM GER-based $N$-best rescoring and system combination work on SPREDS-U1-ja.

\begin{table}[t]
\footnotesize
    \centering
    \caption{Joint $3$ and $4$ systems of Conformer and Whisper (WP) combination checklists with CSJ-trained ASR models. We show the CER [\%] results in original Eval$_{\{1,2,3\}}$ sets in CSJ. }
    \begin{tabular}{cccccc}
    \hline
                 &  WP large v2  &  WP medium  &  Conformer  &  WP small  \\ \hline
    Eval$_1$ (1272) &             3.7    &       4.3        &    5.0      &     5.9        \\ 
    Eval$_2$ (1292) &             2.9    &       3.2        &    3.5      &     4.6        \\ 
    Eval$_3$ (1385) &             3.3    &       3.6        &    4.1      &     5.0        \\ 
    $4$-systems      &      \checkmark    &   \checkmark     &  \checkmark &   \checkmark   \\ 
    $3$-systems      &      \checkmark    &   \checkmark     &  \checkmark &               \\ 
    \hline
    \end{tabular}
    \label{tab:combine_system_checklist}
\end{table}

\begin{table*}[ht] %
\footnotesize
\centering
\caption{Evaluation of LLM GER \red{and MPA GER} by combining $1$-best hypotheses of different systems with CER [\%] on CSJ. \red{Each evaluation data consists of 424 utterances sampled from the original CSJ evaluation data. 
ID means which output was used for MPA GER. 
}}
\begin{tabular}{c|c|c|c|c|c|c|c|c|c}
\hline
  & Whisper &\multicolumn{6}{c|}{$1$-best outputs from $N$-systems } & \multicolumn{2}{c}{Outputs from LLM GER} \\ \cline{3-10}
  & large v2  & \multicolumn{2}{|c|}{LLM GER (Elyza)} & \multicolumn{2}{c|}{LLM GER (Qwen1.5)} & \multicolumn{2}{|c|}{ROVER} & \multicolumn{2}{|c}{MPA GER (proposed)}\\ \cline{3-10}
  &  1-best & $3$-sys & $4$-sys & $3$-sys & $4$-sys & $3$-sys & $4$-sys & $3$-sys & $3$-sys + $4$-sys \\ \hline
ID & A & B1 & B2 & C1 & C2 & D1 & D2 & A+B1+C1 & A+B1+B2+C1+C2 \\ \hline
 Eval$_1$ & 3.4 & 3.3 & 3.4 & 5.0 & 5.1 & 3.5 & 3.6 & 3.4 & \textbf{3.3}\\
 Eval$_2$ & 1.9 & 2.1 & 2.0 & 2.1 & 2.2 & 1.8 & 1.9 & 1.9 & 1.9\\
 Eval$_3$ & 3.0 & 2.9 & 2.9 & 3.1 & 18.8 & 2.9 & 3.1 & 2.9 & \textbf{2.8} \\ 
\hline
\end{tabular}
\label{table:performance_csj_combine_nsystem_1best}
\end{table*}

\begin{table*}[ht] %
\footnotesize
\centering
\caption{Evaluation of LLM GER and MPA GER by combining $3$-best hypotheses of single systems with CER [\%] on CSJ.
\red{Each evaluation data consists of 424 utterances sampled from the original CSJ evaluation data. 
ID means which output was used for MPA GER. }}
\begin{tabular}{l|c|c|c|c|c}
\hline
  &$1$-best  & \multicolumn{3}{c|}{$3$-best outputs from $1$-system} & \multicolumn{1}{c}{Outputs from LLM GER} \\
\cline{3-6}
  & Hypotheses & LLM GER (Elyza) & LLM GER (Qwen1.5) & ROVER & MPA GER (proposed)\\ \hline
ID & A    & B      &  C    & D & A+B+C  \\
\hline
\multicolumn{6}{l}{Whisper large v2}\\
\hline
 Eval$_1$ & 3.4    & 3.7      &  3.9    & 3.7 & 3.5  \\
 Eval$_2$ & 1.9    & 2.3      &  2.2    & 2.2 & 2.0  \\
 Eval$_3$ & 3.0    & 3.3      &  3.5    & 4.0 & 3.1  \\
 \hline
\multicolumn{6}{l}{Whisper medium}\\
\hline
 Eval$_1$ &  3.8   &  4.1     &  4.1    & 4.0 & 3.8  \\
 Eval$_2$ &  2.2   &  2.5     &  2.4    & 2.4 & 2.2  \\
 Eval$_3$ &  3.3   &  3.7     &  3.4    & 4.3 & \textbf{3.2}  \\
 \hline
\multicolumn{6}{l}{Conformer}\\
\hline
 Eval$_1$ &  4.6   & 4.7      &  4.8    & 4.8 & 4.6  \\
 Eval$_2$ &  2.2   & 4.3      &  2.5    & 2.5 & 2.3  \\
 Eval$_3$ &  3.7   & 4.2      &  10.2   & 4.4 & 3.8  \\
 \hline
\multicolumn{6}{l}{Whisper small}\\
\hline
 Eval$_1$ &  5.5   &  5.8     &  5.8    & 5.8 & 5.5  \\
 Eval$_2$ &  3.3   &  3.8     &  5.1    & 3.5 & 3.3  \\
 Eval$_3$ &  4.8   &  4.9     &  4.7    & 5.7 & \textbf{4.7}  \\
 \hline
\end{tabular}
\label{table:performance_csj_1system_nbest}
\end{table*}

\subsection{CSJ Results}
\subsubsection{\red{LLM GER and MPA GER in 1-best $N$-systems}}
Table \ref{table:performance_csj_combine_nsystem_1best} shows the $N$-systems combination LLM GER results on CSJ data. 
Compared to SPREDS-U1-ja, the CER scores on CSJ are much lower.
This would cause a slight improvement with the proposed method on CSJ compared to SPREDS-U1-ja. 
When we compare \red{the LLM GER results between the $3$-systems and $4$-systems, the $4$-systems tends to degrade performance compared to the $3$-systems in Elyza and Qwen 1.5. }
This would be due to the lower performance of Whisper small compared to the other three models. 
In the first step, LLM GER for $3$-systems or $4$-systems, it improved in some cases, but there were various results in different test sets. 
That trend was also seen in direct ROVER results. 
Although the best score of CER=1.8\% in ROVER $3$-systems in Eval$_2$, the score in Eval$_1$ became worse than input scores, and there was no improvement in Eval$_3$. 
In addition, the $4$-systems ROVER also suffered the worst Whisper small model effect and resulted in lower scores than the $3$-systems ROVER. 
We see that the general use of ROVER for some system outputs and the previous LLM GER could not lead to overall improvement, and when some worse systems were included, the harmful effects of such models tended to pull down those models. 
On the other hand, the \red{$3$-systems+$4$-systems in our} proposed MPA GER method resulted in equal or lower overall CERs compared to the inputs. 
In addition, when comparing the proposed MPA GER for \red{$3$-systems+$4$-systems} and $3$-systems, \red{$3$-systems+$4$-systems results} were better than $3$-systems. 
These results support that the LLM combination improved the CER, whereas the other methods tended to worse CER in $4$-systems than $3$-systems.
In the proposed MPA GER, even if a worse system is included, the model can select and incorporate valuable and usable information from the system with a high CER. 

\subsubsection{\red{LLM GER and MPA GER with $N$-best hypotheses in each single system}}
Table \ref{table:performance_csj_1system_nbest} shows the single system $N$-best LLM GER and MPA GER results on CSJ data. 
We applied LLM GER for $3$-best from each single system in Table \ref{tab:combine_system_checklist}. 
We also conducted the experiments in $10$-best settings. 
However, compared to the results in SPREDS-U1-ja with the $10$-best setting for a single system, we could not observe an improvement in the $10$-best system. 
We expect that increasing the size of $N$ in $N$-best for the low CER range ASR task may result in more significant ambiguity of final outputs and result in worse scores.

Overall, the LLM GER ($3$-best) was ineffective and often caused worse scores than $1$-best. 
The proposed \red{MPA GER} was stronger than the \red{traditional combination in ROVER}, and all of the scores exceeded those of ROVER. 
We see that almost ROVER scores were higher than \red{the LLM GER}. 
On the other hand, in the proposed method, we see the lower CER (CER=3.2\%; proposed MPA GER in Whisper medium) or almost the same CER compared to the \red{LLM GER}. 

\begin{table*}[]
\scriptsize
    \centering
    \caption{Example Japanese sentences in different LLM GER and MPA GER systems}
    \begin{tabular}{ll}
    \hline
    \multicolumn{2}{l}{Example1: phonetic similar error (Eval$_1$, $3$-systems+$4$-systems combination)} \\
    \hline
        1-best Hypotheses & \textcolor{purple}{高 速} 条 件 を 導 入 し た 上 で え ー ま 量 子 化 を 行 な っ た 図 な ん で す け ど も ち ょ っ と 量 子 化 の す 話 を 先 に し た い と 思 い ま す  \\
        MPA GER (proposed) & \textbf{\textcolor{blue}{拘 束}} 条 件 を 導 入 し た 上 で え ー ま 量 子 化 を 行 な っ た 図 な ん で す け ど も ち ょ っ と 量 子 化 の す 話 を 先 に し た い と 思 い ま す\\
        Reference & \textbf{\textcolor{blue}{拘 束}} 条 件 を 導 入 し た 上 で え ー ま 量 子 化 を 行 な っ た 図 な ん で す け ど も ち ょ っ と 量 子 化 の す 話 を 先 に し た い と 思 い ま す \\
    \hline    
    \multicolumn{2}{l}{Example 2: phonetic similar error (Eval$_1$, Whisper medium $3$-best) } \\
    \hline
    $1$-best Hypotheses & ミ ュ ン ヘ ン の 博 覧 会 は \underline{電 子 万 華 \textcolor{purple}{経}} の よ う だ と い う 意 味 の 文 で す が え ー 図 の 一 番 上 が \\
    MPA GER (proposed) & ミ ュ ン ヘ ン の 博 覧 会 は \underline{\textbf{電 子 万 華 \textcolor{blue}{鏡}}} の よ う だ と い う 意 味 の 文 で す が え ー 図 の 一 番 上 が \\
    Reference & ミ ュ ン ヘ ン の 博 覧 会 は \underline{\textbf{電 子 万 華 \textcolor{blue}{鏡}}} の よ う だ と い う 意 味 の 文 で す が え ー 図 の 一 番 上 が \\ 	
    \hline
    \multicolumn{2}{l}{Example 3: over-correction for disfluencies (Eval$_1$, Whisper medium $3$-best)} \\
    \hline
    1-best Hypotheses & が 今 回 の え ー っ と 仕 事 の \textcolor{purple}{\underline{え ー っ と}} 大 ま か な も 問 題 意 識 で す ね \\
    MPA GER (proposed) & が 今 回 の え ー っ と 仕 事 の \textcolor{teal}{\underline{え ー と}} 大 ま か な も 問 題 意 識 で す ね \\
    Reference & が 今 回 の え ー と 仕 事 の \textcolor{blue}{\underline{え ー と ー}} 大 ま か な も 問 題 意 識 で す ね \\
    \hline
    \multicolumn{2}{l}{Example 4: \red{repetitions or hallucinations in LLM GER (Eval$_3$, Conformer $3$-best)}} \\
    \hline
    1-best Hypotheses & ... 凄 く 残 念 な ん で す が そ の イ ル ミ ネ ー シ ョ ン を と 大 蔵 山 シ ャ ン ツ と い う \\
    Elyza 3-best  & ... 凄 く 残 念 な ん で す が そ の イ ル ミ ネ ー シ ョ ン を と 大 蔵 山 シ ャ ン ツ と い う \\
    Qwen 1.5 3-best & ... 凄 く 残 念 な ん で す が\textcolor{teal}{そ の イ ル ミ ネ ー シ ョ ン を 見 逃 し て し ま っ て 本 当 見 れ な か っ た の が 凄 く 残 念 な ん で す が...} (repeated in 11 times) \\
    MPA GER (proposed) & ... 凄 く 残 念 な ん で す が そ の イ ル ミ ネ ー シ ョ ン を と 大 蔵 山 シ ャ ン ツ と い う \\
    Reference & ... 凄 く 残 念 な ん で す が そ の イ ル ミ ネ ー シ ョ ン を と 大 倉 山 シ ャ ン ツ ェ と い う\\
    \hline
    \end{tabular}
    \label{tab:sentence_example}
\end{table*}

\begin{figure*}[t]
\centering
  \begin{minipage}[t]{0.45\linewidth}
    \centering
    \includegraphics[width=\linewidth]{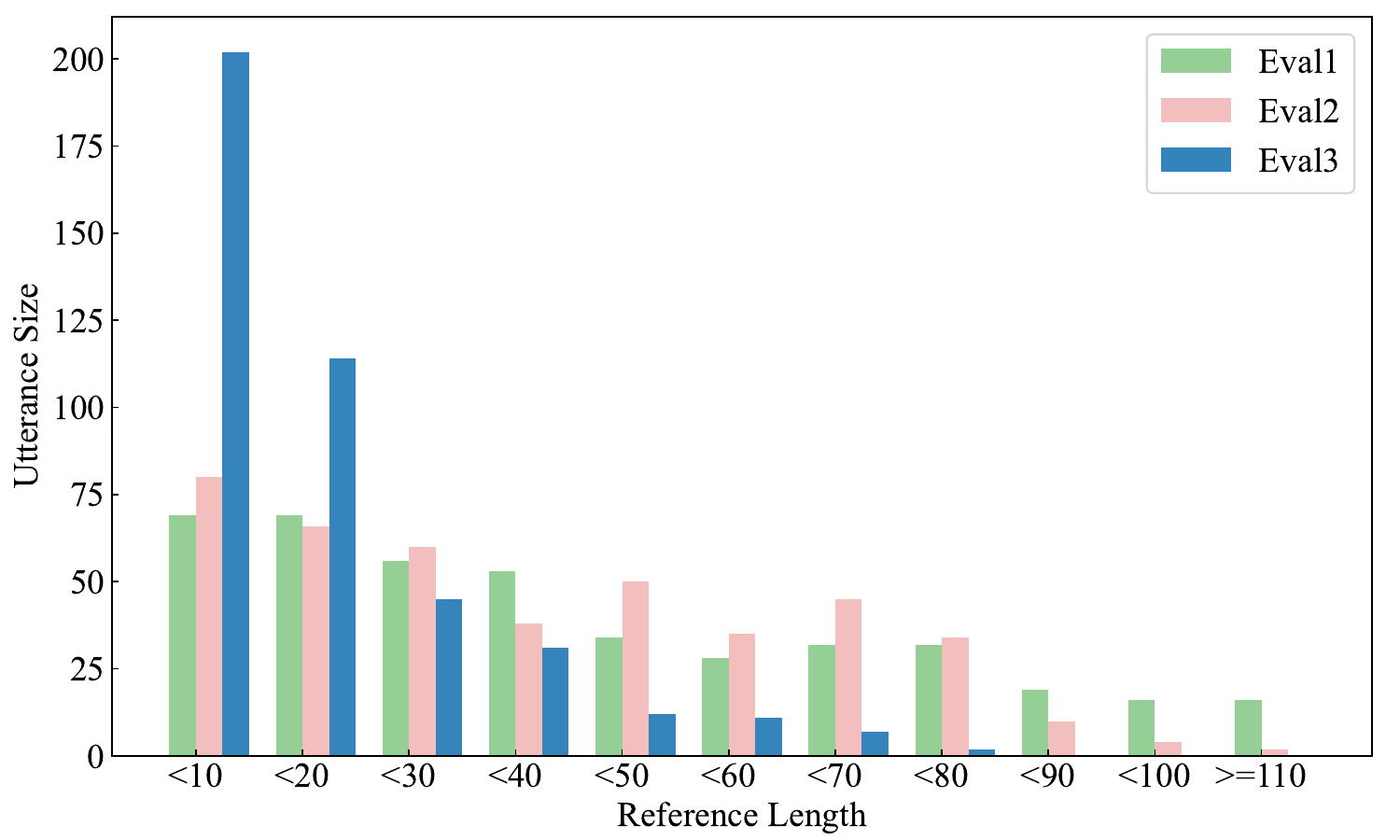}
    \caption{Utterance count by reference length in Eval$_{\{1,2,3\}}$}
    \label{fig:eval1-2-3_reflen_samplenum}
  \end{minipage}
  \begin{minipage}[t]{0.45\linewidth}
    \centering
    \includegraphics[width=\linewidth]{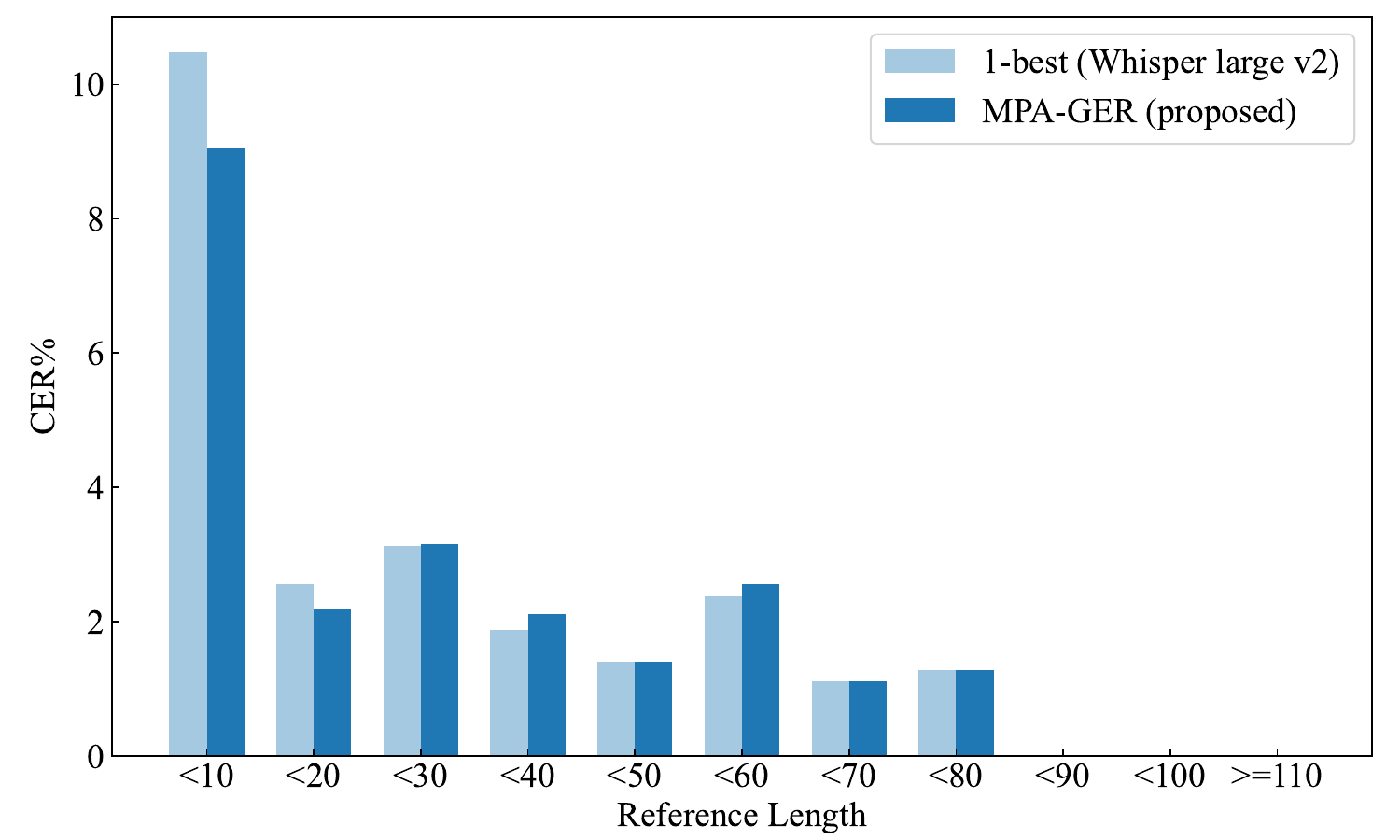}
    \caption{CER [\%] in each reference length in Eval$_3$}
    \label{fig:eval3_reflen_cer}
  \end{minipage}
\end{figure*}

\subsubsection{\red{The GER trends from output examples}}

Table \ref{tab:sentence_example} shows some Japanese correction examples. 
We observed that most fixed errors by the proposed method were categorized as phonetic similarity errors. 
For example, in Example 1 and 2 in Table \ref{tab:sentence_example}, ``\textcolor{purple}{高速}'' was mistakenly generated instead of correct word ``\textcolor{blue}{拘束}'' and ``電子万華\textcolor{purple}{経}'' was mistakenly generated instead of correct word ``電子万華\textcolor{blue}{鏡}'', respectively. 
Those pronunciations were the same in Japanese, and the ASR models generated errors due to such phonetic similarity. 
In our proposed method, the errors were corrected by the proposed MPA GER both in Example 1 and 2. 

Conversely, our methods face challenges when domain-specific context or outer knowledge is necessary for error correction.
We expect that the prompt we used has the restriction to correct errors only with the tokens in the given hypotheses. 
Then, the reconstruction of the prompt effectively injects the outer domain knowledge or grammatical knowledge from the LLM for the GER task. \\
In addition, the model focuses on correcting disfluencies unrelated to the contents of this study. 
We made the LLM models to correct the errors in the outputs, including disfluencies and errors. 
In CSJ data, the domain is spoken language, including disfluencies like fillers in speech and transcripts. 
Example 3 in Table \ref{tab:sentence_example} is over-correcting disfluency parts. 
In Example 3, ``\textcolor{teal}{えーっ}と'' was corrected into ``えーと''. 
However, the model needs to focus on correcting errors related to context words rather than errors related to disfluencies because correcting errors related to disfluencies does not matter for seriously breaking the outputs' meaning.

\subsubsection{Proposed MPA GER alleviates hallucinations or repetitions}
From both Table \ref{table:performance_csj_combine_nsystem_1best} and \ref{table:performance_csj_1system_nbest}, we see the especially worse CERs (CER=18.8\% in Eval$_3$, $3$-systems+$4$-systems, Qwen1.5 in Table \ref{table:performance_csj_combine_nsystem_1best} and CER=10.2\% in Eval$_3$, $3$-best, Qwen1.5, Conformer in Table \ref{table:performance_csj_1system_nbest}).
We observed that the LLM GER error correction generated extended hallucinations or repetitions in those high CER systems. 
Example 4 in Table \ref{tab:sentence_example} shows that the single LLM GER system generates repetitions, but the repetitions were not included in the MPA GER results. 
Such hallucinations or repetitions in the LLM GER are sometimes inevitable and are one reason for degrading performance. 
However, the proposed MPA GER could complement and cancel such hallucinations or repetitions because the tendency of hallucinations or repetitions in each LLM model is different. 
As a result, the proposed MPA GER could get higher performance than ROVER or \red{the LLM GER}, alleviating such \red{hallucination or repetition} effects considering multiple ASR output candidates. 

\subsubsection{Sentence length influences performance}
We analyze whether our error correction is effective for short or long outputs.
Fig. \ref{fig:eval1-2-3_reflen_samplenum} shows the utterance sample numbers in each reference length in Eval$_1$, Eval$_2$ and Eval$_3$. 
We observe that the length distributions are different for Eval$_1$, Eval$_2$, and Eval$_3$. 
In particular, Eval$_1$ and Eval$_2$ are mainly composed of long utterances with a character size of 50 or more, while Eval$_3$ is composed of short utterances with a character size of 10, 20 or less. 
Fig. \ref{fig:eval3_reflen_cer} shows the averaged CERs of 1-best and proposed MPA GER method in each reference length in Eval$_3$. 
From Fig. \ref{fig:eval3_reflen_cer}, we can conclude the reduction trends in CER for short outputs below 10 or 20 and little effect for other relatively long outputs. 
That means the proposed MPA GER is effective, especially for short outputs in this study. 
This led to an overall improvement in the proposed MPA GER, especially in Eval$_3$, which contains more relatively short outputs. 

\section{Conclusion}
\red{In this paper, we applied LLM GER method for Japanese ASR task and introduced a new multi-pass augmented (MPA) GER method that combines hypotheses from different ASR and LLM models. 
From our experiments, we found that combining different LLMs in MPA GER resulted in better CER trends than the previous standard LLM GER results even in low CER ASR task by reducing hallucinations or repetitions in LLMs.
In the future, We will further validate the error correction performance of LLMs in a wider range of ASR scenarios, in a broader range of settings. }

%

\bibliographystyle{IEEEbib}\footnotesize
\bibliography{main_arxiv}
\clearpage
\end{document}